# Towards Open Data for the Citation Content Analysis

© Jose Manuel Barrueco[a], © Thomas Krichel[b], © Sergey Parinov[cb], © Victor Lyapunov[d], © Oxana Medvedeva[b], © Varvara Sergeeva[b]

[a] University of Valencia, Valencia, Spain,
[b] Russian Presidential Academy of National Economy and Public Administration (RANEPA), Moscow, Russia
[c] Central Economics and Mathematics Institute RAS (CEMI RAS), Moscow, Russia
[d] Institute of Computational Mathematics and Mathematical Geophysics SB RAS, Novosibirsk, Russia

**Abstract.** The paper presents first results of the CitEcCyr project funded by RANEPA. The project aims to create a source of open citation data for research papers written in Russian. Compared to existing sources of citation data, CitEcCyr is working to provide the following added values: a) a transparent and distributed architecture of a technology that generates the citation data; b) an openness of all built/used software and created citation data; c) an extended set of citation data sufficient for the citation content analysis; d) services for public control over a quality of the citation data and a citing activity of researchers.

**Keywords:** CitEc, CitEcCyr, open citation data, citation content analysis, digital libraries, RePEc and Socionet

## 1 Introduction

In 1979 Eugene Garfield published his classical book Citation Indexing [1]. Since then data about citations have been used to build bibliometric indicators, such as the impact factor [2] and h-index [3]. The indicators have evaluating research outputs, as well researchers and research institutions that produces these outputs. In 2016, Waltman [4] published a comprehensive literature review on this topic. Exclusive reliance on citation metrics for research evaluation is controversial [5]. However, at the moment all research assessment exercises are heavily based on citation counts.

For many years, the only source of citation data were the indexes published by Institute for Scientific Information. It started as an independent commercial company. Later it was acquired by Thomson Reuters. It is now rebranded as "Clarivate Analytics". Fortunately, the landscape has changed in the last decade. The arrival of alternative sources of citation data has stimulated an increasingly competitive marketplace:

a) In 2004, Elsevier launched Scopus. Today, it is the main competitor of the Web of Science.
b) Scholarly databases, including PubMed or those from Cambridge Scientific, EBSCO and other publishers, are increasingly providing citation information from their core journals.
c) Intermediate services like Mendeley, PlumX or KUDOS are using citation data extracted from Scopus, Web of Science or CrossRef in their products. Social research networks like ResearchGate also generate citations from documents uploaded by researchers.
d) Repositories like arXiv in physics, SSRN in the social sciences and CiteSeer in computer science provide citation counts for papers.

CiteSeer is generally credited with pioneering autonomous citation indexing (henceforth: ACI). ACI is the product of a computer system that automatically creates a citation index from literature available on the internet. It can autonomously locate articles, extract references, identify citations and identify the context of citations in the body of articles [6].

Today, Google Scholar is probably the most popular ACI system. It has the broadest coverage. It is trying to deal with anything available on the web, from any discipline. It has been particularly successful in areas like arts and humanities and in not-English speaking countries, that is, where there were no bibliometric tools before. In Russian research communities, the most popular system is the Russian Citation Index (https://elibrary.ru/project_risc.asp). It is not exactly ACI system, but it is similar. Among Russian researchers and research organizations in economics the ranking systems developed by the IDEAS web site (https://ideas.repec.org/top/) have received a lot of attention. Many of rankings are based on the CitEc (http://citec.repec.org) ACI. However, currently the CitEc data does not cover research papers in Russian.

In 2016, RANEPA (http://www.ranepa.ru/eng/) started a project called CitEcCyr (https://github.com/citeccyr). The project has two main goals: 1) to create a system called CitEcCyr that produces citation relationships for Russian papers and provide these data to the IDEAS ranking system; 2) to create a source of citation data that allows for new types of citation studies, e.g. citation <u>content</u> analysis. All authors of this paper are involved in the CitEcCyr project. As a background, CitEcCyr is using some software and services of CitEc and some data and services at Socionet (https://socionet.ru/).

The remainder of this paper is organized as follows. The second section presents a general view of the CitEcCyr technology. In the third section we discuss specific issues related with upgrading the CitEc tools to process Cyrillic references found in Russian language papers. Section four presents our approach to extracting extended sets of citation data to prepare for citation content analysis. We describe experiments with a new method of conversion PDF documents to JSON data structures and with mining these data structures for sets of citation data. Our concluding section outlines our plan for the future development of the CitEcCyr project.

## 2 A general view of the CitEcCyr technology

We are creating the CitEcCyr technology as an ACI system. To understand how an ACI works, it is critical to grasp the difference between the terms "reference" and "citation" as used within the context of academic papers. A reference found in an academic paper is a description of another academic paper that is mentioned somewhere in the main textual body of the academic paper. References appear outside the main textual body of an academic paper. Most of the time, references appear at the end, in a special section called the reference section. Alternatively, they sometimes appear in footnotes.

When we have a collection of bibliographic data about academic papers, then each of the papers in that collection will have a handle that identifies the paper. If we find that a paper *A* contains a reference that is a description of paper *B*, we say that paper *A* cites paper *B*. If we can find the handle for the paper *B* we can build a citation relationship between papers *A* and *B*. The citation relationship is a data record that contains the handles of the citing paper and the cited paper.

In the last twenty years, there has been a sustained movement for open access to academic papers. It has received much attention. Building open-access metadata collections about papers has received much less attention. But it is still an importance concern. To put it in simple terms, a collection of PDF files on a disk or a set of web location does not make for a digital library, just like pile of books is not a physical library. Hiding in plain sight away from the attention that open access has received, people have assembled digital libraries of papers. RePEc and Socionet are very fine examples of such digital libraries. Since we require some level of standardization among digital libraries involved in our current work, we define a <u>Com</u>mon <u>Res</u>earch <u>I</u>nformation <u>S</u>pace (henceforth: ComRIS), which is a combination of compatible parts of RePEc and Socionet.

In Figure 1, the ComRIS and its services, used in CitEcCyr project, are on the left side.

To build an ACI, we need reference lists from the papers. Sometimes, publishers provide references as a supplementary data set. See the box "Paper has references in XML/TEXT" at the Figure 1. If we don't have reference lists from publishers, we may use the full-text of documents to get to reference lists. Let's call that process "reference detection". It is a cumbersome task. If a full text is in PDF we try to extract its content. This is the box "Conversion PDF to txt/json" at the Figure 1. From there we still need to get the citation data, including references and the mentioning in the main textual body. We discuss these issues in details in the section 4.

Once we have references, we can parse them to find data that we use to build citation relationships. It is the box "Parse references" at Figure 1.

Finally we find what references refer to what papers. This is the box of the "Reference linking" at the Figure 1. To detect a citation, our software uses three data elements from a reference: (1) the author string, (2) the title, and (3) the publication year. Thus, these are the fields we have to parse the reference for.

CitEc, as a parent system for CitEcCyr, does not use its own software to do either of the reference detection or the reference parsing work. It relies on ParsCit [7]. It uses this software "out of the box", that is, without any customization. This is possible for two reasons. First, ParsCit is cleverly designed to run an any Unix system basically from the console. And second, ParsCit was built by computer scientist. Fortunately, computer science and English-language economics enjoy very similar citation styles. However, for the Russian references that we have in the ComRIS, we found ParsCit be working very poorly. Following user complaints, CitEc currently excludes Russian documents. So, in CitEcCyr we have to do a Russian references processing service almost from the scratch.

As the Figure 1 demonstrates, each step of the CitEcCyr data flow is based on the output of the previous one. In order to pass a certain step, the document has to pass all previous ones. If it does not, the document is marked as unprocessed and an error code is assigned to it.

## 3 From CitEc to CitEcCyr

The CitEcCyr system is using the CitEc architecture and much of its software as a prototype. At this time, CitEc and CitEcCyr are two separate systems. The interaction between them still has to be fully developed.

### 3.1 Data

At the outset we neither have a reference detection nor a reference parsing system for Russian papers available. But we have some reference data from the Neicon archive[1] at Socionet. This set of references contains 1014099 strings at the time of writing. 535576 of these reference strings do not contain any Cyrillic

---

characters at all.

Now this is already raises an interesting question. When does a reference really become a Russian one? The question transits from being interesting to being vexing when one realizes that the dataset contains both transliterated and translated references. There are cases when translations and transliterations are mixed in with a Cyrillic references.

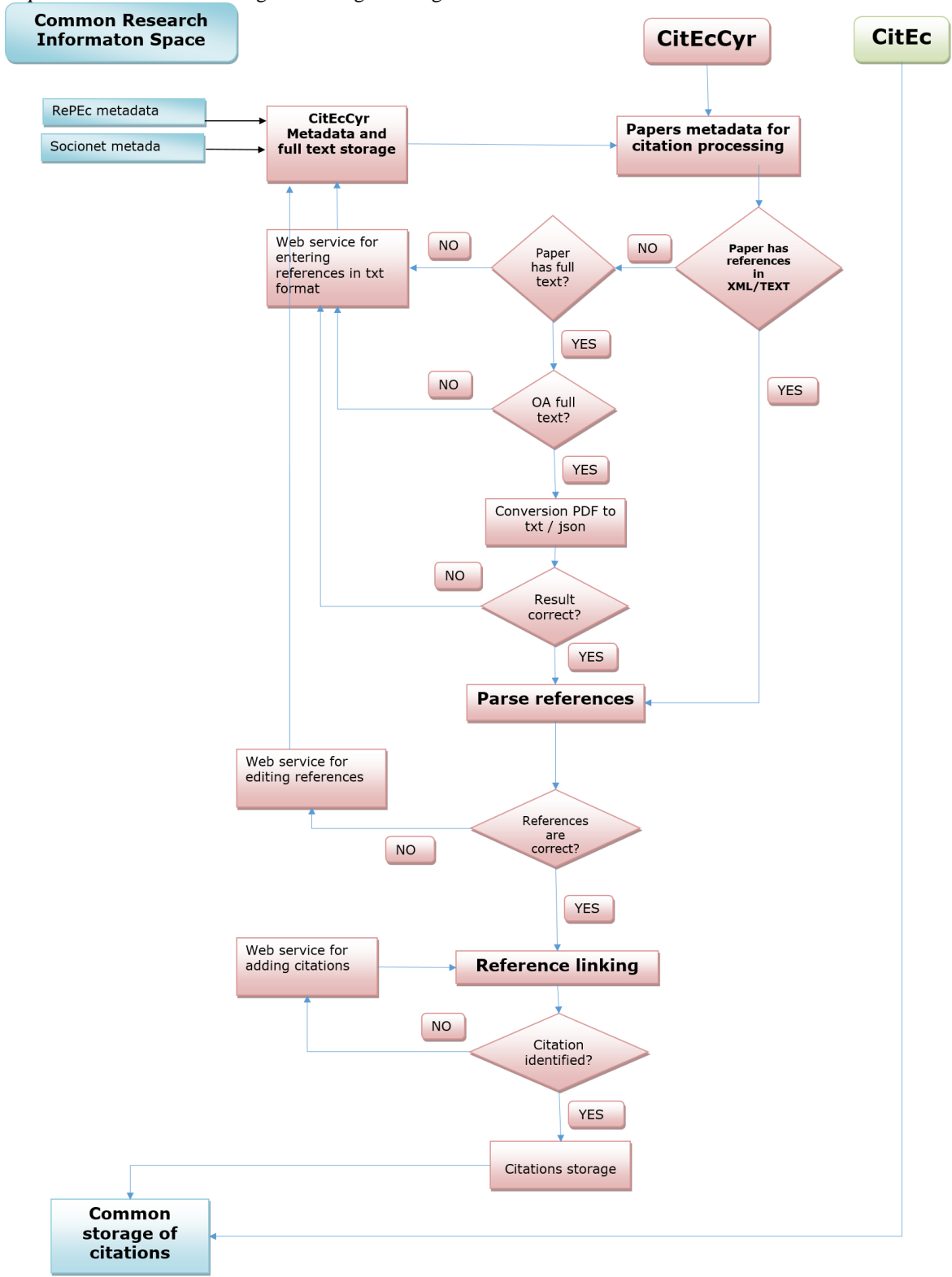

Fig. 1. CitEcCyr data flow diagram

Thus the reference parsing suddenly appears to become a two-stage problem. First detect where a Cyrillic part is. Then parse the Cyrillic part[2]. But detecting the Cyrillic part is not as easy as it seems.

Sometimes the data features homoglyph errors. By definition, homoglyphs are characters that look very much the same. We use the term here to mean that they look identical. It does turn out that dataset is quite heavily polluted with such homoglyphs, for example where the Latin "e" is used instead of the Cyrillic "е".

Another problem with the data is over- or underblanking. "Overblanking" occurs when there are additional blanks between letters like in "Д а ш е в с к и й В.Я., К а ц н е л ь с о н А.М.". Underblanking occurs when blanks are missing like in "Майстренко Н.А., Шейко С.Б., Алентьев А.В. и сотр.//Практическая онкология. -2008. -Т. 9, № 4. -С. 229-236." Here, tokenziation at blanks would make the author field to include "//Практическая". Thus tokenization at blanks is problematic.

### 3.2 Software

Since CitEc has been using ParsCit successfully, even though not on Russian references, it appears logical to want to use it for CitEcCyr. It may appear that it would just be a matter of extending it. Unfortunately, it is not so easy.

ParsCit is based on a statistical learning approach known as "conditional random fields" (henceforth: coraf). The idea is to look at a sequence of fields. A field is a sequence of tokens that have any characteristic we want to learn something about. The important thing is to understand that fields appear in sequence. The sequence may not be regular but the idea is that the inclusion of a token in one field will impact the distribution of likely values in the next one.

As a learning technology, coraf needs a training dataset. This dataset has essentially to be manually prepared. The ParsCit training dataset comes from computer science papers. Interestingly enough, computer science and economics use similar reference formats. Thus there is no need to compose a separate training dataset.

ParsCit uses a set of dictionaries for author names, place names, publishers. They can be replaced, but one has to replace the handling software as well as it has bugs.

The feature generation part of ParsCit is not modularized. You could add or remove features. But allowing for different feature building codes to coexist would require major refactoring. Thus after careful consideration, we decided to build a separate set of code that would replace the reference parsing functionality that ParsCit provides. We would not use ParsCit for this. We still use the crf++ coraf software of 工藤 拓 (Tako Kudo), available at https://taku910.github.io/crfpp/. It is the same software that ParsCit uses. But we use it in a different way.

### 3.3 Tokenizing

One decision that affects everything else and that has to be resolved first is how reference lines are to be tokenized.

Fields will be collections of tokens. No field border can be drawn within a token. Tokenizing at white-space is the default choice. We thought that tokenizing at word borders would be better. It is only when we have that finer granularity that we would be able to deal with underblanked lines. However, upon thinking a long time, we now believe that word-border tokenization would bring too many cumbersome problems when punctuation accumulates. It would also create many spurious tokens like, for example, in a URL.

Thus we stick to a basic white-space tokenization.

### 3.4 Learning data

We wrote bespoke software to prepare learning data. This is done by using a custom XML format, constrained by a RelaxNG specification. It's best understood through an example. We pick a line: *Горденко Э.А. Варлаам Хутынский и архиепископ Антоний в житиях и мистериях XII-XVI века. – М.; СПб., 2010.*

Recall that we only need to parse author names titles and the publication year. The person who prepares the training data writes an XML fragment:
*<r><a>Горденко Э.А.</a> <t>Варлаам Хутынский и архиепископ Антоний в житиях и мистериях XII-XVI века.</t> – М.; СПб., <y>2010.</y></r>*

In fact that fragment is prepopulated with the text from the reference dataset. Other software translates the fragment into

| | |
|---|---|
| Горденко | B-A |
| Э.А. | I-A |
| Варлаам | B-T |
| Хутынский | I-T |
| и | I-T |
| архиепископ | I-T |
| Антоний | I-T |
| в | I-T |
| житиях | I-T |
| и | I-T |
| мистериях | I-T |
| XII-XVI | I-T |
| века. | I-T |
| – | O |
| М.; | O |
| СПб., | O |
| 2010. | B-Y |

Here "B-" means begin, "I-" means inside. "-A" stands for the author field, "-T" stands for the title field and "-Y" stands for the year field. An "O" in the second column suggests that the token is not in any field. The software has to be able to translate it back into a string as well to check whether or not there has been some corruption.

---

in Belorussian, and maybe Ukrainian.

### 3.5 Dictionaries

We have mentioned that ParsCit has dictionaries. These are simple lists of terms. We have compiled two for our Cyrillic reference parsing software.

The first is for names. We extracted 14944 Russian family names by copying manually from http://gufo.me/fam_a. This is a seed dataset. Since there are male and female names, we can add derived female names for male ones and vice-versa. Then, purpose-written software combs the reference lines for occurrences of the family names, followed by initials. If we find a similar construct, meaning a capitalized word followed by some upper case letters, followed by a known author name, we have a new candidate name. A name of the opposite sex may be derived, and then the scanning resumes. Over time, this procedure has gathered over 100000 names.

The second is for abbreviations. In general, the most difficult aspect of reference parsing is to find the end of the title. In the presence of the double slash, it stops at the double slash. In the absence of a double slash, the best indicator is the first word with a period at the end. If, however, the dotted term is an abbreviation, we need to skip it or mark it as such. Thus, we use a bespoke list of abbreviations that are most common in the reference lines.

### 2.6. Features

This is where the magic comes in. Most of the strings seem to orient themselves after ГОСТ 7.1-2003[3], or ГОСТ Р 7.0.-2008[4]. The data that we have is generally a sequence of author family names with initials, title, and then a tail with everything else. In that tail, we also find personal names, but they will be editor names that we don't count for reference linking.

Conceptually, we group features into two types. We call them "token features" and "row features". A token feature can be determined from the individual token. A row feature is based on all values that the token features took for all the tokens in the reference lines. Thus row features are evaluated after the token features.

The first token features in the broad type of a token. It can be Cyrillic, a number, a URL, a symbol, a Roman numeral, or something undetermined. Generally, when faced with mixed Latin and Cyrillic characters, the algorithm tries to figure out if changing homoglyphs will convert to Latin and Cyrillic only. Otherwise, we have settle for an undetermined broad type. The following features are all binary. The second are known author family names. The third are capitalized words. The fourth is all upper cases with optional dots. Then we have a publication year. Then we test for a continuous four digits. We also use special indicators for the mimeo indicator "M.:", the double slash and the single slash.

After this follow the row features. They evaluate the entire row. The two are indicators whether a token is before or after the single slash or a double slash, respectively. There is feature that indicates a likely inclusion in the author field. Finally there is an indication of whether a token appears to be in the title.

With such elaborate features, the main role of the coraf software is to add as an aggregator of sophisticated, purpose-build features.

### 3.7 Results

The procedure classifies tokens into "A" for author, "T" for title, "Y" for year and "O" for other. The crf++ coraf software provides us with estimates of its own accuracy at the level of each token.

Table 1. Self-assessment of quality of parsing for fields.

| field | count | mean | var. | min | max |
|-------|----------|------|------|------|------|
| A | 6586212 | 0.77 | 0.04 | 0.30 | 0.99 |
| O | 59134372 | 0.93 | 0.01 | 0.26 | 0.99 |
| T | 33731520 | 0.79 | 0.01 | 0.24 | 0.97 |
| Y | 377869 | 0.86 | 0.01 | 0.33 | 0.97 |

### 3.8 Conclusion

These first results suggests that is possible to parse ГОСТ-inspired citations with sophisticated features without sophisticated use of coraf.

## 4 Open Data for Citation Content Analysis

This section is about achieving the second goal of the CitEcCyr project. That is to create an open source of citation data allowing new types of citation studies, e.g. citation content analysis. There are two popular ways to denote the incident when a reference is mentioned in a paper text[5]. These are "in-text citation" and "in-text reference", respectively. In this paper, we use the second term. It is the key data for citation content analysis

### 4.1 Citation Content Analysis concept

The authors of the Citation Content Analysis (CCA) framework wrote [8]: "CCA is mainly established on two rationales: 1) instead of being weighted equally citations should be granted different weights under different contexts; 2) qualitative measurements (e.g., how one cites) and quantitative measurements (e.g., number of citations) should be incorporated and mutually complementary."

To make this possible we need data on how cited papers are mentioned in the text of citing papers. One task is to recognize in a paper all in-text references and link them unequivocally with appropriate reference data listed at end of a paper. A context around of the in-text reference should also give us useful data for the CCA's "qualitative measurements".

Some part of the Table 4. "Two-dimension and two-modular code book for CCA" from the paper [8] give us ideas about possible indicators of the citation content

analysis:
- Location of mentioning: 1) Abstract; 2) Introduction; 3) Literature Review; 4) Methodology; 5) Results/discussion; 6) Conclusion; 7) Others.
- Frequency of mentioning: 1) Once; 2) 2 to 4 times; 3) 5 times or more.
- Style of mentioning: 1) Not specifically mentioning; 2) Specifically mentioning but interpreting; 3) Direct quotation.

To determine the "location of mentioning" we have to find the titles of all the sections in a paper and map them according some classification ("introduction", "review", "methodology", etc.). We also have to determine text coordinates for each section in a paper, since we should build relationships between the sections and in-text references extracted from a paper.

If we parse the in-text references correctly, there is no problem to count a frequency of mentioning for each in-text reference.

To determine the style of mentioning we have to have a text located on the left and on the right for each in-text reference. When we analysis this text, we can classify in-text references.

The first step is a proper conversion method to create text versions of binary PDF documents.

### 4.2 PDF conversion

We used the PDF.js[6] as a processor to create a structured text from PDF documents. We use the Node.js platform[7] to run this Java-script module server-side. Our conversion package is an open source software available at GitHub[8]. The result is a very efficient conversion procedure. On average, it processes a PDF document in 1-2 seconds.

The output is a JSON version of PDF document. It is instructive to look at the Example 1 with a small fragment of a JSON file produced from a PDF document[9].

Example 1. JSON version of PDF document, a fragment

```
…{"page":1,"textContent":{"items":[
{"str":" ","dir":"ltr","width":1.2,
"height":23.04,"transform":[4.8,0,0,4.8,118.319
9,736.64],"fontName":"g_d0_f1"},
{"str":"Available online at ",
"dir":"ltr","width":75.63096,"height":78.854399
99999998,"transform":[8.879999999999999,0,0,8.8
79999999999999,182.9868,788.48],"fontName":"g_d
0_f2"},
{"str":"www.sciencedirect.com","dir":"ltr","wid
th":94.038312,"height":78.85439999999998,"trans
form":[8.879999999999999,0,0,8.879999999999999,
258.5376,788.48],"fontName":"g_d0_f2"}…
```

A full version of this JSON file is available[10].

The text of the document is in the `"str"` attribute. Other attributes report on formatting. These attributes include `"page"`, `"transform"`, `"fontName"` and others. These attributes give data for computer recognition within the paper's content: section titles, headers and footers, page numbers, comparative position of words on a document's page, etc. The JSON version of a PDF document allows for counting of text coordinates. These in turn, which can be used to visualize the PDF documents for readers. Since PDF.js is used by Hyopthes.is, the in-text references can be visualized as annotations. Such visualization allows: (a) to present citation statistics and other results of CCA over PDF content; (b) to provide transparency for results of CCA; (c) to enable for public control over correctness of CCA data.

To test this approach, we made PDF to JSON conversion for PDF papers from the Neicon collections we mentioned in section 3.1. This archive currently includes 150 collections of research papers with about 65000 papers in total. At the beginning of June 2017 we processed a random sample of about 10000 of PDF documents. The results of the conversion are available at https://socionet.ru/~cyrcitec/json/spz/neicon/. They include initial PDF documents and its JSON versions. As file names we used handles of the same papers at Socionet.

Only one PDF document of the processed set had a processing error. This PDF file has a broken PDF structure. Some JSON versions of PDF documents have no text, because the PDF documents have no text layer.

### 4.3 Parsing citation data

Using JSON versions of PDF papers we are parsing extended citation data, including:
a) the content of lists of references, which we need to get to the citations;
b) in-text references, which give us data for counting CCA indicator the "Frequency of mentioning";
c) text context around the in-text references, which allow specify CCA's "Style of mentioning".

CCA also requires to parse the section titles and find their text coordinates. This is necessary to specify the location of in-text references, to make its weighting by the "Location of mentioning" specified in CCA. This specific data parsing will be implemented at the next stage of our project.

Below we provide examples of results of parsing extended citation data. They are based on the JSON file from Example 1. All extracted citation data is stored as XML files.

Example 2 provides extracted data about a reference. This data includes following attributes:
a) the raw text of a reference, see it below in the tag ``;
b) the position of the reference in the reference list, see attribute num;
c) the text coordinates of the reference, see the attributes start and end;
d) the attributes author, title and year

---

extracted from data in the tag `` and using as input data for linking the reference with metadata of the same paper.

Example 2.Parsing a reference data

```
…
<reference num="4" start="27513" end="277
80" author="Parinov S. " title="Towards
an Open Data on how the Research Data are
Used CRIS CERIF based Approach"
year="2014">
        Parinov S. Towards an
        Open Data on how the Research Data
        are Used: CRIS CERIF based
        Approach. In the proceedings of the
        12th International Conference on
        Current Research Information
        Systems (CRIS 2014).
        2014
</reference>
…
```

Full extracted data about references are available[11].

Example 3 illustrates extracted data about an in-text reference and its context. We are parsing here only one type of in-text references marked as a reference number in square brackets. The data we extract includes:

a) a number of the in-text reference, the tag `<Reference>`;

b) symbols of the in-text reference, tag `<Exact>`;

c) text coordinates of the in-text reference, tags `<Start>` and `<End>`;

d) a context located at the left, the tag `<Prefix>`, and at the right, the tag `<Suffix>`, according the in-text reference.

To parse a context we just take 200 symbols before and after the string in the tag `<Exact>`.

Example 3.Parsing an in-text reference data

```
…
<intextref>
        <Reference>4</Reference>
        <Exact>[4]</Exact>
        <Start>3950</Start>
        <End>3952</End>
        <Prefix>forms of the research
        outputs usage by integrating the
        semantic linkage technique into
        CRIS functionality [3],</Prefix>
        <Suffix>. As a result, a pilot of
        the open semantically enrichable
        research information system for
        researchers [5] has been
        provid</Suffix>
</intextref>
…
```

Full extracted data of in-text references is available[12].

The citation data parsed from 10000 JSON files of the Neicon archive papers are available at http://no-xml.socionet.ru/~cyrcitec/citmap/spz/neicon/.

### 4.4 Linking

As described in the section 2 we are linking references with metadata of the same papers, if the papers are available at Socionet. If the linking act is successful, we include the handle - see the attribute "`handle`" in the Example 4 - of the paper's metadata into the XML files with parsed citation data.

Example 4. Linked reference data

```
<reference num="4" start="27605" end="278
31" author="Parinov" title="Towards an
Open Data on how the Research Data are
Used CRIS-CERIF based
Approach" year="2014"
handle="RePEc:rus:mqijxk:34">
```

Additionally, we are managing a list of unlinked references. If for some reference we can't find the cited paper, we check whether we can find the reference at the list of unlinked references. If we can't we add this reference to the list. If we find it, we create a temporary handle from the unlinked reference list. We include this handle into the XML date - see the attribute "`handle`" in the Example 5 - the same as we do this for regular handles.

Example 5. Unlinked reference data

```
<reference
        num="8"
        start="28684"
        end="28819"
        url="https://dx.doi.org/10.6084/m9.
figshare.871466.v1"
        author="Preston Johnston"
        title="The Future of Academic
Research"
        year="2013"
        handle="spz:cyrcitec:references:3">
```

Using the list of unlinked references we have a handle for each unlinked reference. Now we can use them as regular information object. It allows:

a) processing them at Socionet like citation relationships, i.e. counting of a number of citations for unlinked references, building of citation hyperlinks between citing and cited papers, etc.;

b) simplification of processing the arrival at Socionet of new papers that match some unlinked references, and then moving them into the set of linked references;

c) providing authors of papers with a web interface to correct errors in their unlinked references to move the references into the set of linked references.

For the cases "b" and "c" above the procedure of "moving" unlinked references into the set of linked ones implies that we replace the temporal handle of unlinked reference by the handle of a paper recently appeared at Socionet that matches the reference.

As a result, we update XML tags with parsed references data to replace in their `handle` attribute temporal handles by the regular ones.

## 5 Conclusion

In the middle of June 2017 the CitEcCyr project provides for public:

1) a big set of JSON versions of PDF documents, which can be used for parsing different data from research papers;

2) a big set of XML files with the first results of parsing citation data, which can be used for different types of citation content analysis.

The next steps of the project are:

a) to run regular processing of citation relationships from RePEc papers in Russian to include these data into IDEAS ranking system;

b) to created regularly updated open source of citation data sufficient for citation content analysis.

## Acknowledgments

We are grateful to Alexei Skalaban for helping us with the Neicon data. We are grateful to Min-Yen Kan for answering questions regarding ParsCit.